\documentstyle[preprint,aps]{revtex}
\begin{document}

\draft

\title{ZGB model with random distribution of inert sites}

\author{ \sc { G.L. Hoenicke and W. \ Figueiredo }} 
\address{Departamento de F\'{\i}sica - Universidade Federal de Santa
Catarina\\ 
88040-900, Florian\'opolis, SC  - Brasil; e-mail:
wagner@fisica.ufsc.br, hoenicke@fisica.ufsc.br }

\maketitle

\begin{abstract}
 A random distribution of inert sites is introduced in  the  Ziff-Gulari-Barshad model to study the phase transitions 
between active and  poisoned states. The adsorption of ${\rm CO}$ and ${\rm O_2}$ molecules is not possible at the position  
of the inert sites. This model is investigated in  the site and  pair approximations, as well as  through Monte Carlo 
simulations. We determine  the mean coverages of the elements as a function of the dilution and    show that the continuous 
transition between the active  and  O-poisoned state is  slightly affected by moderate values of dilution  in the pair 
approximation and in the simulations. On the other hand, from the analysis of the hysteresis curves,  the transition between the  active and CO-poisoned states  changes 
from first-order to continuous as one  increases   the concentration of inactive sites. The observed transition in the site and pair approximations is always of first-order nature. We also found the lines of transition and spinodal points as a function of the concentration of inert sites. Finally,  the production rate of ${\rm CO_2}$ is calculated as a function of the dilution of sites. 
\end{abstract}

PACS number(s): {05.70.Ln, 05.70.Fh, 82.65.Jv, 82.20.Mj}     

\newpage
\vskip2pc

\section{Introduction}

   The study of nonequilibrium phase transitions is a topic of growing interest due to its application to a variety of 
complex systems$^{1,2}$: contact process, domain growth, catalysis, phase separation and transport phenomena. Although 
there is no general theory to account for nonequilibrium model systems, in recent years some progress has been  
achieved in understanding the stationary states of these systems employing approximate analytical methods and 
simulations. Some rigorous mathematical questions concerning the phase transitions of these complex  interacting 
particle systems can be appreciated in the books of Liggett$^{3}$ and Konno$^{4}$.

In this paper we focus our attention  on the phase transitions observed in the surface reaction model proposed by 
Ziff, Gulari and Barshad$^{5}$ (ZGB), which  describes some kinetic aspects of the oxidation of ${\rm CO }$ over  a catalytic 
surface. In particular, here  we consider  a modified version of the ZGB model, where we incorporate a random 
distribution of inert sites on the catalytic surface. The original  ZGB model is an   irreversible lattice model 
for surface reactions based on the Langmuir-Hinshelwood mechanism, where the reactants must be adsorbed before 
reacting. The  steps used  to describe the  ZGB model ( a lattice Markov process) are the following: Molecules 
of ${\rm CO}$ and ${\rm O_2}$ from a gaseous phase can be adsorbed onto the sites of a regular square lattice of identical 
sites. These molecules arrive at the surface according to their partial pressures in the gas mixture, that is, 
the probability of a  ${\rm CO}$ molecule arriving  is  $y_{{\rm co}}$ and $(1-y_{{\rm co}})$ for the ${\rm O_2}$ molecule. The ${\rm CO}$ 
molecule requires only a single vacant site to be adsorbed, while the ${\rm O_2}$ is adsorbed if it finds  a 
nearest-neighbor pair of empty sites. Upon adsorption, the   ${\rm O_2}$ molecule  dissociates and the two free  ${\rm O}$ atoms   
can react independently. If,  after an adsorption step, a  nearest-neighbor ${\rm CO-O}$ pair appears on the lattice, 
they immediately react, forming a ${\rm CO_2}$ molecule that  goes to the gas phase, leaving two empty sites on the 
lattice. Therefore, in this adsorption controlled limit, only a single parameter ($y_{{\rm co}}$) is sufficient to 
describe the dynamics of the model.

The simulations performed by Ziff and co-workers have  shown that the system exhibits two phase transitions between 
active and poisoned states: for $y_{{\rm co}} \leq y_1$, an O-poisoned state is formed, while for $y_{{\rm co}} \geq y_2$ 
the lattice is poisoned by ${\rm CO}$. For $y_1 < y_{{\rm co}} < y_2$  a reactive steady-state is found, in which  a nonzero 
number of vacant sites is present in the lattice. At $y_1$ the transition is continuous, whereas  at $y_2$ the 
transition is of the first-order type.  Using a mean field theory, Dickman$^{6}$ qualitatively  reproduced the phase diagram 
of the ZGB model and showed that, at the level of site approximation,  only the first-order transition appears. However, employing the 
pair approximation, both continuous and first-order transitions are obtained.

 We  are interested  on the effects of inert sites on the phase transitions of the ZGB model.  We have 
investigated  in detail the dependence of the phase transitions on  the concentration of inert sites. This 
problem presents some experimental interest in the automobile industry, where   lead particles are 
deposited over the catalyst   during the exhaust of the gases after combustion. This affects the 
efficiency of the catalytic  surface  due to  the pinning of these lead particles on the  surface,   forbidding 
the adsorption of ${\rm CO}$ and ${\rm O_2}$ molecules at the lead positions and reducing the  reaction paths. Hovi 
and co-workers$^{7}$,  have studied by computer simulations the effect of preadsorbed poison and promoters 
on the irreversible ZGB model. They calculated the coverage of species as a function of the concentration 
of inert sites for a wide range of values, finding  the interesting result that the first-order transition 
changes to a continuous one at  a critical  value of the concentration.  Cort\'es and Valencia$^{8}$ have also reported some results concerning random impurities distributed over the catalyst, in which  they observed the change of the first-order transition into a continuous one as one  increases  the concentration of impurities. Albano$^{9}$ simulated the ZGB model on  incipient percolation clusters (IPC's) with a fractal dimension of 1.90. He showed that both transitions,  at $y_1$ and $y_2$  are continuous,  and that for an infinite lattice, in which  $y_{{\rm co}}$ is  larger than 0.408, the reactions stop at finite times because the IPC's are  poisoned by pure ${\rm CO}$. Casties et al.$^{10}$ also performed a Monte Carlo simulation of the ${\rm CO}$ oxidation on probabilistic fractals. They observed a change in the character of the transition at $y_2$ from first order on regular lattices to second order on percolation clusters (for $p$ larger than $p_c=0.593$,  which is  the percolation threshold  on the square lattice).

	In this work we have performed  mean-field ( site and pair approximations) calculations  and Monte 
Carlo simulations for different values of the concentration of inert sites. The model studied here is a variant 
of the original ZGB model, where inert sites are randomly distributed over the lattice. Our  approach is  close related to that  presented by  Vigil and Willmore$^{11}$ to study  the effects of spatial correlations on the oscillatory behavior of a modified  ZGB model,   where defects are continually added and desorbed from the surface. In their studies, they considered the mean-field site  and pair approximations, as well as Monte Carlo  simulations.  In the present  work we have determined  the phase 
diagram for different  concentrations, and the spinodal and transition lines  as a function of the concentration of inert 
sites. We have constructed hysteresis curves to find the critical concentration  at which the first-order 
transition changes into a continuous one. This paper is organized as follows: in Sec. II we present the results 
obtained within the site approximation; in  Sec. III  we introduce the pair approximation equations and   
show  the results obtained using this scheme; Sec. IV presents the results of simulations, and finally, in 
Sec. V we present our conclusions.

\section{Site approximation}
 
	We take a square lattice as our catalytic  surface. A fraction $n_d$ of the sites is randomly distributed 
over the lattice  representing  the pinned inert sites. The remaining sites of the lattice can be vacant, or  occupied 
by either  ${\rm O}$ atoms or ${\rm CO}$ molecules. The ZGB model is described by the following steps:
\begin{eqnarray}
{\rm CO(g)} + v \rightarrow {\rm CO(a)}, \\
{\rm O_2(g)}+ 2v \rightarrow 2{\rm O(a)},\\
{\rm CO(a)} + {\rm O(a)} \rightarrow {\rm CO_2(g)} + 2v, 
\end{eqnarray}
where the labels $g$ and $a$ denote the  gaseous phase and an adsorbed reactant on the surface, respectively, and 
$v$ indicates a vacant site. Steps (1) and (2) indicate the adsorption of the species, whereas the third step 
is the proper   reaction,  between  distinct species located at adjacent sites of the lattice. In the site 
approximation the time evolution equations  of the concentrations are given by
\begin{eqnarray}
\frac{d{\rm  n_o}}{dt} &=&-y_{{\rm co}}n_v(1-(1-{\rm  n_o})^4)+2(1-y_{{\rm co}})n_v^2(1-n_{{\rm co}})^3 , \\
\frac{dn_{{\rm co}}}{dt} &=&y_{{\rm co}}n_v(1-{\rm  n_o})^4-2(1-y_{{\rm co}})n_v^2(1-(1-n_{{\rm co}})^3) ,
\end{eqnarray}
where  ${\rm  n_o}$, $n_{{\rm co}}$ and $n_v$ represent, respectively, the coverages of ${\rm O}$, ${\rm CO}$ and blank sites in the lattice. 
$y_{{\rm co}}$ gives the arrival probability of  a ${\rm CO}$ molecule. In addition, there is the following constraint among the concentrations
\begin{equation}
n_{{\rm co}}+ {\rm  n_o} + n_v + n_d =1 .
\end{equation}

The steady-state solutions of the above system of equations  are given by $n_v = 0$, that corresponds to a poisoned surface, and
\begin{eqnarray}
n_v &=& \frac {y_{{\rm co}}}{2(1-y_{{\rm co}})}= Y .
\end{eqnarray}
Inserting Eq. (7) into Eq. (4) we obtain an expression for the steady-state values of the concentration $n_{{\rm co}}$:
\begin{eqnarray}
 (n_{{\rm co}}+Y+n_d)^4+(1-n_{{\rm co}})^3-1 = 0. 
\end{eqnarray}

We exhibit in Fig. 1 a typical  diagram for the coverages of ${\rm CO}$, ${\rm O}$ and vacant sites obtained for $n_d=0.2$. This  diagram was obtained by integrating  
  the equations of motion for the $n_{{\rm co}}$ and ${\rm  n_o}$ concentrations,  starting  from an initial condition in which  the 
number of empty sites is $n_v=1-n_d$. The site approximation does not give any continuous transition for all values 
of the concentration of inert sites. This was already pointed out   by Dickman$^{6}$ for the ZGB model without inert sites. 
We observe in Fig. 1, that the limit of stability of the reactive phase is    $y_s=0.467510$, which  corresponds to the 
spinodal point. Therefore,  a reactive steady-state is found  for all values of $y_{{\rm co}} \leq y_s$. For values of 
$y_{{\rm co}} > y_s$, the system becomes poisoned, with a large amount of ${\rm CO}$ and a small  concentration of ${\rm O}$ atoms. 
The presence of ${\rm O}$ atoms in the region $y_{{\rm co}} > y_s$ is due to the inert sites that can  block some oxygen,  
and also to the simplicity of the site approximation,  which  does not forbid the formation of  ${\rm O-CO}$ nearest-neighbor
 pairs in the lattice. The tolerance of these ${\rm O-CO}$ pairs also explains the absence of the  continuous phase transition, which is observed in   the simulations. Fig. 2 is a plot of the solutions $n_{{\rm co}}$  of Eq.(8) versus the parameter $Y$ for 
different values of concentration $n_d$ of inert sites. We obtain two solutions, which we call $n^{>}_{{\rm co}}$ and 
$n^{<}_{{\rm co}}$, that join together at the spinodal point. For instance, for $n_d=0$ the value we find is $Y=0.638986$, 
which  furnishes the value $y_s=0.561013$. We also note in Fig. 2 that,  at the spinodal point, the concentration of 
$n_{{\rm co}}$ molecules remains the same irrespective of the  value we choose for $n_d$. This special value is $n_{{\rm co}}=0.1660$.  Then,  the net effect of adding $n_d$ is to shift the curves horizontally. 
In this site approximation,   solutions are possible only for values of $n_d < 0.638986$. This happens  because above 
this value the solution would correspond to the non-physical value $y_{{\rm co}} < 0$. So, the meaning of the  two solutions in 
Fig. 2 is the following: the branch $n^{<}_{{\rm co}}$ represents  the stable steady-state solutions whereas   the $n^{>}_{{\rm co}}$ branch  gives  the unstable solutions. These solutions were obtained after numerical integration of the equations of motion for $n_{{\rm co}}$ and ${\rm  n_o}$,  starting from the state described by   $n_v=1-n_d$.  For the initial condition $n_v=Y$ and  $n_{{\rm co}}$ larger than 
 $n^{>}_{{\rm co}}$ the system evolves to the poisoned state. The initial condition $n_v=Y$ and $n_{{\rm co}}$ less than 
$n^{>}_{{\rm co}}$ drives the system to the lower stable reactive  solution $n^{<}_{{\rm co}}$. 

Fig. 2 also shows that, as we approach the spinodal point for any value of $n_d$, the region of stability becomes 
narrower. Then, we expect that for some value of $y_{{\rm co}} \leq y_s$ a first-order transition occurs, that is, the  
concentration $n_{{\rm co}}$ must increase abruptally from a small value ( reactive phase) to a large value (poisoned phase). 
Unfortunately,  we cannot  use here the usual thermodynamic considerations based on the minimization of a suitable 
thermodynamic potential. In order to find this first-order transition we adopt the same kinetic criteria employed 
by Dickman$^{6}$, which was borrowed from the work of Ziff et al$^{5}$. The phase transition was determined by 
choosing an initial state where half of the lattice was empty and the other half was completely filled with ${\rm CO}$. 
In this work we choose as our initial state,  to solve the equations of motion for $n_{{\rm co}}$ and ${\rm  n_o}$,  the  values 
$n_v = n_{{\rm co}}= \frac{1}{2}(1-n_d)$. It is clear that this choice is not the same as that  considered  by Ziff et al., because we 
cannot  discriminate which sites are empty or not. The phase boundary is defined at the special value $y_2$ where 
the solution of the equations of motion changes from the reactive to the poisoned state as we vary the value of 
$y_{{\rm co}}$ for the same  initial condition,  as established above.  For $n_d=0$ we obtain the same value found by Dickman. 
We exhibit in Fig. 3 the results obtained for the first-order transition and the spinodal points for $n_d=0.2$. The spinodal 
was obtained from the initial condition $n_v=1-n_d$,  and the   first-order transition from the condition 
$n_v = n_{{\rm co}}= \frac{1}{2}(1-n_d)$. For this particular value of $n_d=0.2$, we have $y_2= 0.3999$ and $y_s=0.4821$. 
We have considered all values of the  concentration of inert sites, and Fig. 4 shows the values of $y_s$ (dashed line) 
and $y_2$ (full line) as a function of the concentration of inert sites. At the particular value $n_d=0.55$ the 
two lines merge. For  values of $n_d > 0.55$ the transition still remains  of the  first-order type,  although the number 
of vacant sites that  stay   in the active state  is very small. For instance, for $n_d=0.60$, at the transition point ($y_2=0.0725$),
the number of vacant sites changes from $0.0388$ in the active state, to $ 2\times 10^{-8}$ in the poisoned state.  Throughout  our analysis we considered a given state to be active if the number of vacant 
sites is larger  than $ 10^{-6}$. We also exhibit in Fig. 5  the number of vacant sites $n_v$ at the active state as 
a function of the number of inert sites $n_d$, at the transition and spinodal points. We observe that for all values 
of $n_d < 0.55$, the number of vacant sites at the spinodal point is always larger  than that  at the transition point.

\section{Pair approximation}

Let us consider the application of the pair approximation procedure to this ZGB model that includes  inert sites. 
Here we introduce the pair probability $P_{\alpha \beta}$ of  a random nearest neighbor pair of sites being  
occupied by species $\alpha$ and $\beta$. We have the following types of species: $v$, $d$, $c$, and ${\rm O}$, 
which represent, respectively, vacant, inert, carbon monoxide, and oxygen. As in the previous treatments 
$^{12,13}$ we need to consider only  the changes that occur at a particular central pair in the lattice. 
In the table below we display  the allowed and forbbiden (indicated by $\times$) nearest-neighbor pairs 
in the present model.
\[
\begin{tabular}{|c|c|c|c|c|}
\hline
$\quad\quad$ & $v$ & $o$ & $c$ & $\,\,\,\,d\,\,\,\,\,$ \\ \hline
$v$ & $vv$ & $vo$ & $vc$ & $vd$ \\ \hline
$o$ & $\quad\quad $ & $oo$ & $\times $ & $od$ \\ \hline
$c$ & $\quad\quad $ & $\times $ & $cc$ & $cd$ \\ \hline
$d$ & $\quad\quad $ & $\quad\quad $ & $\quad\quad $ & $dd$ \\ \hline
\end{tabular}
\]

The next table   also  exhibits  all the possible transitions among pairs. We obtain $14$ independent  transitions, 
labelled by numbers in the range $1-14$. In the table  transitions indicated by $\times$ are prohibited.
\[
\begin{tabular}{|c|c|c|c|c|c|c|c|c|}
\hline
$\frac{\text{from }\longrightarrow }{\text{to }\downarrow }$ & $vv$ & $vo$ & 
$vc$ & $vd$ & $oo$ & $od$ & $cc$ & $cd$  \\ \hline
$vv$ & $\quad\quad $ & $4$ & $6$ & $\times $ & $\times $ & $\times $ & $12$ & $\times $
 \\ \hline
$vo$ & $1$ & $\quad\quad $ & $\times $ & $\times $ & $10$ & $\times $ & $\times $ & $%
\times $  \\ \hline
$vc$ & $2$ & $\times $ & $\quad\quad $ & $\times $ & $\times $ & $\times $ & $13$ & $%
\times $  \\ \hline
$vd$ & $\times $ & $\times $ & $\times $ & $\quad\quad $ & $\times $ & $11$ & $\times $
& $14$  \\ \hline
$oo$ & $3$ & $5$ & $\times $ & $\times $ & $\quad\quad $ & $\times $ & $\times $ & $%
\times $  \\ \hline
$od$ & $\times $ & $\times $ & $\times $ & $8$ & $\times $ & $\quad\quad $ & $\times $
& $\times $  \\ \hline
$cc$ & $\times $ & $\times $ & $7$ & $\times $ & $\times $ & $\times $ & $\quad\quad $
& $\times $ \\ \hline
$cd$ & $\times $ & $\times $ & $\times $ & $9$ & $\times $ & $\times $ & $%
\times $ & $\quad\quad $  \\ \hline 
\end{tabular}
\]

Then, we write the   equations relating the probability of each element with the corresponding pair probabilities:
\begin{equation}
\begin{array}{c}
P_v=P_{vv}+P_{vo}+P_{vc}+P_{vd}, \\ 
P_o=P_{od}+P_{vo}+P_{oo}\hspace{0.43in}, \\ 
P_c=P_{cd}+P_{vc}+P_{cc}\hspace{0.45in}, \\ 
P_d=P_{dd}+P_{vd}+P_{od}+P_{cd}.
\end{array}
\end{equation}

The pair probabilities also  satisfy  the constraint 
\begin{eqnarray}
P_{vv}+P_{oo}+P_{cc}+P_{dd}+2(P_{vo}+P_{vc}+P_{vd}+P_{od}+P_{cd})=1.
\end{eqnarray}

Next,  we need to write the time evolution equations for the pair probabilities.  Examining   the latter table we can  
construct the desired equations of evolution. We explicitly write the equations of motion for the pair probabilities $P_{\alpha\beta}$.
\begin{eqnarray}
\frac{dP_{vo}}{dt} &=&t_1+t_{10}-t_4-t_5, \nonumber \\ 
\frac{dP_{vc}}{dt} &=&t_2+t_{13}-t_6-t_7, \nonumber \\ 
\frac{dP_{vd}}{dt} &=&t_{11}+t_{14}-t_8-t_9, \nonumber \\ 
\frac{dP_{oo}}{dt} &=&t_3+2t_5-2t_{10}, \nonumber  \\ 
\frac{dP_{od}}{dt} &=&t_8-t_{11}, \\ 
\frac{dP_{cc}}{dt} &=&2t_7-t_{12}-2t_{13}, \nonumber\\ 
\frac{dP_{cd}}{dt} &=&t_9-t_{14}, \nonumber \\ 
\frac{dP_{vv}}{dt} &=&-t_3+t_{12}-2t_1+2t_4-2t_2+2t_6, \nonumber\\
\frac{dP_{dd}}{dt} &=&0. \nonumber 
\end{eqnarray}
where $t_1$ to $t_{14}$ are the transition rates. The factors of two arising in the equation of motion for $P_{vv}$ 
are due to  the fact that the pair probabilities  $P_{ij}$ and $P_{ji}$ are equal by symmetry. For instance, from the 
pair $vv$ we can obtain,  with  the same  probability, the different configurations $ov$ and $vo$. In general, the 
expressions for the transition rates are lengthy, and we present these  transition rates in the Appendix.

	In this pair approximation we cannot obtain analytical solutions as we have   done in the site approximation. We 
solved the coupled set of eight nonlinear equations by the fourth-order Runge-Kutta method, searching for the 
stationary solutions. We considered the two different  initial conditions as in the case of the site approximation.

	Let us first  consider the evolution from the initial state where $P_v=1-P_d$, in which  only the pairs   $vv$, 
$vd$ and $dd$ are present in the lattice at $t=0$. Figure 6 shows the   diagram of the model for $P_d=0.2$. 
For $0 < y_{{\rm co}} \leq 0.2299$ the lattice poisons with oxygen. In the range $0.2299 < y_{{\rm co}} < 0.4821$ there is  an 
active region, and for $y_{{\rm co}} \geq 0.4821$ the lattice poisons  with ${\rm CO}$. When $P_d=0$, we found  the same figures  
obtained by Dickman in his pair approximation. For instance, the site and pair approximations give the same value 
for the spinodal point $y_s$. However, when we consider some inert sites in the lattice, the spinodal point found in   the site approximation is always smaller  than that obtained within the pair approximation. For this particular 
value,  $P_d=0.2$, the  site approximation yields  $y_s=0.4675$, whereas   $y_s=0.4821$ is obtained by the pair approximation.  The value of 
$y_{{\rm co}}$  at the continuous transition,  which now  arises  in this pair approximation,   decreases slightly with increasing 
values of the  concentration of inert sites.

	We also considered the solutions evolving  from an initial condition where half of the free sites  ($P_v=1-P_d$) is 
filled with ${\rm CO}$ molecules and the other half left  empty. In order to be close to the  initial condition used in   the simulation,  
we chose for the initial  pair conditions     $P_{cc}=P_{vv}$, $P_{dv}=P_{dc}$ and $P_{vc}=0$, which mimics a division of 
the lattice  into two parts: on  one side of the lattice  we would have inert sites and ${\rm CO}$ molecules  and,   on the other 
side,  vacant and inert sites. If $P_d=0$, we found for the transition between the  active and CO-poisoned states   the value 
$y_2=0.5240$, which agrees with the value found in the simulations. Fig. 7 displays the concentration of ${\rm CO}$ molecules 
at the transition point for which  $P_d=0.2$. In this pair approximation, the values of $y_s$ and $y_2$ are very close. We 
also show in Fig. 8 the concentration of vacant sites as a function of the concentration of inert sites, at the 
transition point, and also at the spinodal point. Both curves join at $P_d=0.50$, and  for $P_d > 0.60$, we cannot  
observe any active state. As in the site approximation, an active state is defined only if $P_v > 10^{-6}$. Then, 
the calculations performed within the  pair approximation give results that are very similar to those obtained by the  site approximation,  concerning 
the spinodal and transition points.

	In addition, it was   observed that initial conditions do not affect the point in  which   the continuous phase transition occurs.  
In Fig. 9 we exhibit the phase diagram for this ZGB model with inert sites. The size of the reactive window decreases 
as we increase the concentration of inert sites. We have  plotted the transition  line for the first-order transition 
and for  the spinodal line,  which gives the limit of stability of the reactive phase. The line separating the active and 
O-poisoned phases is a continuous transition line.

\section{Simulations}

        We have performed Monte Carlo simulations  in the ZGB model with inert sites in order to check the results 
we have obtained in the site and pair approximations. The simulations were carried for different values of the 
concentration of inert sites $P_d$. For small values of $P_d$, we considered square lattices of linear size $L=64$, 
but for large values of $P_d$ we have  used 
lattices of  linear size up to $L=150$. The first step in the simulation is to  randomly distribute the selected 
fraction $P_d$ of inert sites in the lattice. All simulations then started with a fraction  of empty sites equal 
to $P_v=1-P_d$. The ${\rm CO}$ molecules arrive at the surface with a probability $y_{{\rm co}}$ and the ${\rm O_2}$ molecules with 
probability $1-y_{{\rm co}}$. The rules for adsorption and reaction of the species are exactly the same  as in the original ZGB 
model$^{5}$. Since  adsorption of oxygen requires two nearest neighbor empty  sites, the effect of the inert sites 
is to favour the adsorption of ${\rm CO}$ relatively to that of  ${\rm O_2}$ molecules. In general,  we have taken $10^{3}$ 
Monte Carlo steps (MCs) to attain the stationary states, and  $10^{3}$ more  to calculate the concentration 
averages at the stationary states. One MCs is equal to $L \times L$ trials of deposition of species, where 
$L$ is the linear size of the lattice. To speed up the simulations we worked with a  suitable list of empty sites.

	We exhibit in Fig. 10 the phase diagram of the model in the plane $y_{{\rm co}}$ versus $P_d$. It is  
similar to that  obtained within the  pair approximation. However,  there is a fundamental  difference between 
the transition line separating the active and CO-poisoned phases in both approaches. In the pair approximation 
the transition line is always of the first-order type, whereas   in the simulations there is a critical concentration 
above which the transition becomes  continuous. We have done detailed simulations to find the critical concentration 
at which  the transition becomes continuous. We have found  for the critical concentration of inert sites the 
value $P^{c}_{d}=0.078$. We   arrived at this value by looking at the  hysteresis loops in the curves of $P_{{\rm co}}$ 
versus $y_{{\rm co}}$ for different values of the concentration $P_d$, as we can see in Fig. 11. We proceed as follows:  
in Fig. 11a we fixed the concentration of inert sites at the value $0.070$ and the curve with circles, which 
is the proper transition curve, was obtained   from  an  initial state where $P_v=1-P_d$, that is, with 
a lattice almost empty. The curve  with squares was determined from an initial state in which  the lattice was 
almost covered by ${\rm CO}$.  We have taken a fraction of only $5\%$ of randomly empty sites over the lattice at 
the starting time. Then,  we clearly observe the hysteresis loop at the  concentration $P_d$, which implies 
that  the transition is of first-order. On the other hand, Fig. 11b, where the fraction of inert sites is 
$P_d=0.080$, does not exhibit the hysteresis loop  and the transition is clearly a continuous one. The 
critical value of $P^{c}_{d}=0.078$ was obtained analysing the behavior of these curves in the range 
$0.070 < P_d < 0.080$. As we have pointed out in the Introduction,  Hovi et al.$^{7}$ had already  observed  the change in the nature of this  transition as a function of the  concentration.  The phase boundary separating the active and the O-poisoned phases in Fig. 10 is 
continuous for all values of $P_d$. We have checked this fact by observing that no hysteresis loop was 
found for any value of $P_d$. The width of the active phase decreases with increasing values of $P_d$. 
For values of $P_d > 0.45$ the lattice is poisoned (absence of empty sites) with different amounts of 
${\rm CO}$ and ${\rm O}$ species. Due to  finite size effects, this value is larger than the value  0.408  found  by Albano$^{9}$ in the limit of very large IPC's.

	We have also noted that the production rate of ${\rm CO_2}$ molecules attains its maximum value exactly at 
the first-order transition,   for values of $P_d < P^{c}_{d}$. If $P_d > P^{c}_{d}$ the maximum production rate 
of ${\rm CO_2}$ molecules is located inside of the reactive window. This is seen  in Fig. 12, where the   circles indicate the points where the production rate of ${\rm CO_2}$ is maximum.  In the site and pair 
approximations this maximum occurs always at the phase boundary, irrespective of the value  of  $P_d$.  Fig. 13 shows  the production rate $R$ of ${\rm CO_2}$ molecules as a function of $P_d$. As   expected, the role 
of inert sites is also of   blocking the reactions over the catalyst. The maximum production rate occurs 
at a surface free of impurities.

\section{Conclusions}

   We have studied the effects of a random distribution of inert sites on the phase diagram of the ZGB model. 
We determined the time evolution equations for the concentrations of the different species over the catalytic 
surface within an effective field theory, at the level of  site and  pair approximations, and  also performed 
Monte Carlo simulations on the model. We obtained  the coverages of the  species as function of the deposition rate of 
${\rm CO}$ and of the  concentration of inert sites. In the site and pair approximations we found the transition line  and 
the limit of stability of the reactive phase. In the site approximation, the continuous transition between 
the O-poisoned and reactive states is absent for any values  of the concentration of inert sites. The width of 
the reactive window exhibits the same behavior,  as a function of concentration of inert sites,  in both pair 
approximation and Monte Carlo simulations. However, the transition  between the reactive and CO-poisoned phase 
is always of first-order in the site and  pair approximations, whereas  Monte Carlo simulations  give  a critical 
point where the transition changes nature. For values of the concentration of inert sites less than the 
critical value,  the transition is first-order and above this value, it changes to a continuous one. The 
determination of this critical concentration was possible through  the analysis of the  hysteresis curves for 
different values of the  concentration of inert sites. The production rate of ${\rm CO_2}$ molecules is maximum at the 
first-order transition,  in both site and pair approximations. This is the case in  the simulations, but  the transition 
is of the first-order type.  When the concentration of inert sites is greater than the critical value,  the maximum production 
rate of ${\rm CO_2}$ molecules moves towards the reactive window. The overall effect of inert sites is to reduce the 
production of ${\rm CO_2}$ molecules.

\section*{Acknowledgments}

We would like to thank    Ron Dickman by  his many valuable  suggestions, and  Luis G. C. Rego  by  the critical reading of the manuscript.  This work was supported by the Brazilian agencies CAPES, CNPq and FINEP.

\appendix

\section*{Transition rates in  the pair approximation}

We present the transition rates in  the pair approximation, which we used in  Section III to solve the time 
evolution equations for the pair probabilities. The transition rates are $t_1$ to $t_{14}$, which are given by
\begin{eqnarray}
t_1&=&t_{1a}+t_{1b} \\
t_{1a}&=&(1-y_{co})P_{vv}3\frac{P_{vv}}{P_v}\left( 1-\frac{P_{vc}}{P_v}\right) ^2 
\nonumber \\
t_{1b}&=&(1-y_{co})P_{vv}\left( 1-\frac{P_{vc}}{P_v}\right) ^3\left\{ 1-\left( 1-%
\frac{P_{vc}}{P_v}\right) ^3\right\}   \nonumber
\end{eqnarray}

\begin{equation}
t_2=y_{co}P_{vv}\left( 1-\frac{P_{vo}}{P_v}\right) ^3
\end{equation}

\begin{equation}
t_3=(1-y_{co})P_{vv}\left( 1-\frac{P_{vc}}{P_v}\right) ^6
\end{equation}

\begin{eqnarray}
t_4 &=&t_{4a}+t_{4b} \\
t_{4a} &=&y_{co}P_{vo}\left\{ \left( 1-\frac{P_{vo}}{P_v}\right) ^3+\frac 32%
\frac{P_{vo}}{P_v}\left( 1-\frac{P_{vo}}{P_v}\right) ^2+\right.  \nonumber \\
&&\left. \left( \frac{P_{vo}}{P_v}\right) ^2\left( 1-\frac{P_{vo}}{P_v}%
\right) \frac 14\left( \frac{P_{vo}}{P_v}\right) ^3\right\}  \nonumber \\
t_{4b} &=&y_{co}P_{vo}\left\{ 3\frac{P_{vo}}{P_o}\left[ \left( 1-\frac{P_{vo}}{P_v%
}\right) ^3+\frac 32\frac{P_{vo}}{P_v}\left( 1-\frac{P_{vo}}{P_v}\right)
^2+\right. \right.  \nonumber \\
&&\left. \left. \left( \frac{P_{vo}}{P_v}\right) ^2\left( 1-\frac{P_{vo}}{P_v%
}\right) +\frac 14\left( \frac{P_{vo}}{P_v}\right) ^3\right] \right\} 
\nonumber
\end{eqnarray}

\begin{equation}
t_5=(1-y_{co})P_{vo}3\frac{P_{vv}}{P_v}\left( 1-\frac{P_{vc}}{P_v}\right) ^2
\end{equation}

\begin{eqnarray}
t_6 &=&t_{6a}+t_{6b} \\
t_{6a} &=&(1-y_{co})P_{vc}3\frac{P_{vv}}{P_v}\left\{ \left( 1-\frac{P_{vc}}{P_v}%
\right) ^2+\frac{P_{vc}}{P_v}\left( 1-\frac{P_{vc}}{P_v}\right) +\frac 13%
\left( \frac{P_{vc}}{P_v}\right) ^2\right\}  \nonumber \\
t_{6b} &=&3(1-y_{co})P_{vc}\frac{P_{vc}}{P_c}\left\{ 3\frac{P_{vv}}{P_v}\left[
\left( 1-\frac{P_{vc}}{P_v}\right) ^2+\frac{P_{vc}}{P_v}\left( 1-\frac{P_{vc}%
}{P_v}\right) +\frac 13\left( \frac{P_{vc}}{P_v}\right) ^2\right] \right\} 
\nonumber
\end{eqnarray}

\begin{equation}
t_7=y_{co}P_{vc}\left( 1-\frac{P_{vo}}{P_v}\right) ^3
\end{equation}

\begin{equation}
t_8=(1-y_{co})P_{vd}3\frac{P_{vv}}{P_v}\left( 1-\frac{P_{vc}}{P_v}\right) ^2
\end{equation}

\begin{equation}
t_9=y_{co}P_{vd}\left( 1-\frac{P_{vo}}{P_v}\right) ^3
\end{equation}

\begin{eqnarray}
t_{10} &=&t_{10a}+t_{10b} \\
t_{10a} &=&y_{co}P_{oo}\frac{P_{vo}}{P_o}\left\{ \left( 1-\frac{P_{vo}}{P_v}%
\right) ^3+\frac 23\frac{P_{vo}}{P_v}\left( 1-\frac{P_{vo}}{P_v}\right)
^2+\left( \frac{P_{vo}}{P_v}\right) ^2\left( 1-\frac{P_{vo}}{P_v}\right)
\right.  \nonumber \\
&&+\left. \frac 14\left( \frac{P_{vo}}{P_v}\right) ^3\right\}  \nonumber \\
t_{10b} &=&2y_{co}P_{oo}\frac{P_{vo}}{P_o}\left\{ \left( 1-\frac{P_{vo}}{P_v}%
\right) ^2\left( 1-\frac{P_{oo}}{P_o}\right) +\frac 12\left[ \frac{P_{oo}}{%
P_o}\left( 1-\frac{P_{vo}}{P_v}\right) ^2\right. \right.  \nonumber \\
&&+\left. 2\frac{P_{vo}}{P_v}\left( 1-\frac{P_{vo}}{P_v}\right) \left( 1-%
\frac{P_{oo}}{P_o}\right) \right] +\frac 13\left( \frac{P_{vo}}{P_v}\right)
^2\left( 1-\frac{P_{oo}}{P_o}\right)  \nonumber \\
&&+\left. \frac 23\frac{P_{vo}}{P_v}\left( 1-\frac{P_{vo}}{P_v}\right) \frac{%
P_{oo}}{P_o}+\frac 14\left( \frac{P_{vo}}{P_v}\right) ^2\frac{P_{oo}}{P_o}%
\right\}  \nonumber
\end{eqnarray}

\begin{eqnarray}
t_{11} &=&t_{11a}+t_{11b} \\
t_{11a} &=&y_{co}P_{od}\frac{P_{vo}}{P_o}\left\{ \left( 1-\frac{P_{vo}}{P_v}%
\right) ^3+\frac 32\frac{P_{vo}}{P_v}\left( 1-\frac{P_{vo}}{P_v}\right)
^2+\left( \frac{P_{vo}}{P_v}\right) ^2\left( 1-\frac{P_{vo}}{P_v}\right)
\right.  \nonumber \\
&&+\left. \frac 14\left( \frac{P_{vo}}{P_v}\right) ^3\right\}  \nonumber \\
t_{11b} &=&2y_{co}P_{od}\frac{P_{vo}}{P_o}\left\{ \left( 1-\frac{P_{vo}}{P_v}%
\right) ^2\left( 1-\frac{P_{od}}{P_d}\right) +\frac 12\left( 1-\frac{P_{vo}}{%
P_v}\right) ^2\frac{P_{od}}{P_d}\right.  \nonumber \\
&&+\frac{P_{vo}}{P_v}\left( 1-\frac{P_{vo}}{P_v}\right) \left( 1-\frac{P_{od}%
}{P_d}\right) +\frac 13\left( \frac{P_{vo}}{P_v}\right) ^2\left( 1-\frac{%
P_{od}}{P_d}\right)  \nonumber \\
&&+\left. \frac 23\frac{P_{vo}}{P_v}\left( 1-\frac{P_{vo}}{P_v}\right) \frac{%
P_{od}}{P_d}+\frac 14\left( \frac{P_{vo}}{P_v}\right) ^2\frac{P_{od}}{P_d}%
\right\}  \nonumber
\end{eqnarray}

\begin{eqnarray}
t_{12} &=&2(1-y_{co})P_{cc}\left( \frac{P_{vc}}{P_c}\right) ^2\left\{ \left( 1-%
\frac{P_{vc}}{P_v}\right) ^4+2\left( 1-\frac{P_{vc}}{P_v}\right) ^3\frac{%
P_{vc}}{P_v}\right.  \nonumber \\
&&+\frac 23\left( \frac{P_{vc}}{P_v}\right) ^2\left( 1-\frac{P_{vc}}{P_v}%
\right) ^2+\left( \frac{P_{vc}}{P_v}\right) ^2\left( 1-\frac{P_{vc}}{P_v}%
\right) ^2  \nonumber \\
&&+\left. \frac 23\left( \frac{P_{vc}}{P_v}\right) ^3\left( 1-\frac{P_{vc}}{%
P_v}\right) +\frac 19\left( \frac{P_{vc}}{P_v}\right) ^4\right\}
\end{eqnarray}

\begin{eqnarray}
t_{13} &=&t_{13a}+t_{13b} \\
t_{13a} &=&(1-y_{co})P_{cc}\frac{P_{vc}}{P_c}3\frac{P_{vv}}{P_v}\left\{ \left( 1-%
\frac{P_{vc}}{P_v}\right) ^2+\frac{P_{vc}}{P_v}\left( 1-\frac{P_{vc}}{P_v}%
\right) +\frac 13\left( \frac{P_{vc}}{P_v}\right) ^2\right\}   \nonumber \\
t_{13b} &=&2(1-y_{co})P_{cc}\frac{P_{vc}}{P_c}\left\{ 2\frac{P_{vv}}{P_v}\left[
\left( 1-\frac{P_{vc}}{P_v}\right) \left( 1-\frac{P_{cc}}{P_c}\right) +\frac 
12\frac{P_{vc}}{P_v}\left( 1-\frac{P_{cc}}{P_c}\right) \right. \right.  
\nonumber \\
&&+\left. \frac 12\left( 1-\frac{P_{vc}}{P_v}\right) \frac{P_{cc}}{P_c}+%
\frac 13\frac{P_{vc}}{P_v}\frac{P_{cc}}{P_c}\right] +\frac{P_{vc}}{P_c}%
\left[ \left( 1-\frac{P_{vc}}{P_v}\right) ^2\right.   \nonumber \\
&&+\left. \frac{P_{vc}}{P_v}\left( 1-\frac{P_{vc}}{P_v}\right) +\frac 13%
\left( \frac{P_{vc}}{P_v}\right) ^2\right] \times \left. \left[ \frac{P_{vc}%
}{P_v}\left( 1-\frac{P_{vc}}{P_v}\right) +\frac 23\left( \frac{P_{vc}}{P_v}%
\right) ^2\right] \right\}   \nonumber
\end{eqnarray}

\begin{eqnarray}
t_{14} &=&t_{14a}+t_{14b} \\
t_{14a} &=&(1-y_{co})P_{cd}\frac{P_{vc}}{P_c}\left\{ 3\frac{P_{vv}}{P_v}\left[
\left( 1-\frac{P_{vc}}{P_v}\right) ^2+\frac{P_{vc}}{P_v}\left( 1-\frac{P_{vc}%
}{P_v}\right) +\frac 13\left( \frac{P_{vc}}{P_v}\right) ^2\right] \right\}  
\nonumber \\
t_{14b} &=&2(1-y_{co})P_{cd}\frac{P_{vc}}{P_c}\left\{ 2\frac{P_{vv}}{P_v}\left[
\left( 1-\frac{P_{vc}}{P_v}\right) \left( 1-\frac{P_{cd}}{P_d}\right) +\frac 
12\frac{P_{vc}}{P_v}\left( 1-\frac{P_{cd}}{P_d}\right) \right. \right.  
\nonumber \\
&&+\left. \frac 12\left( 1-\frac{P_{vc}}{P_v}\right) \frac{P_{cd}}{P_d}+%
\frac 13\frac{P_{vc}}{P_v}\frac{P_{cd}}{P_d}\right] +\frac{P_{vd}}{P_d}%
\left( 1-\frac{P_{vc}}{P_v}\right) ^2+\frac 22\frac{P_{vc}}{P_v}\frac{P_{vd}%
}{P_d}  \nonumber \\
&&+\left. \left. \frac 13\left( \frac{P_{vc}}{P_v}\right) ^2\right] \right\} 
\nonumber
\end{eqnarray}

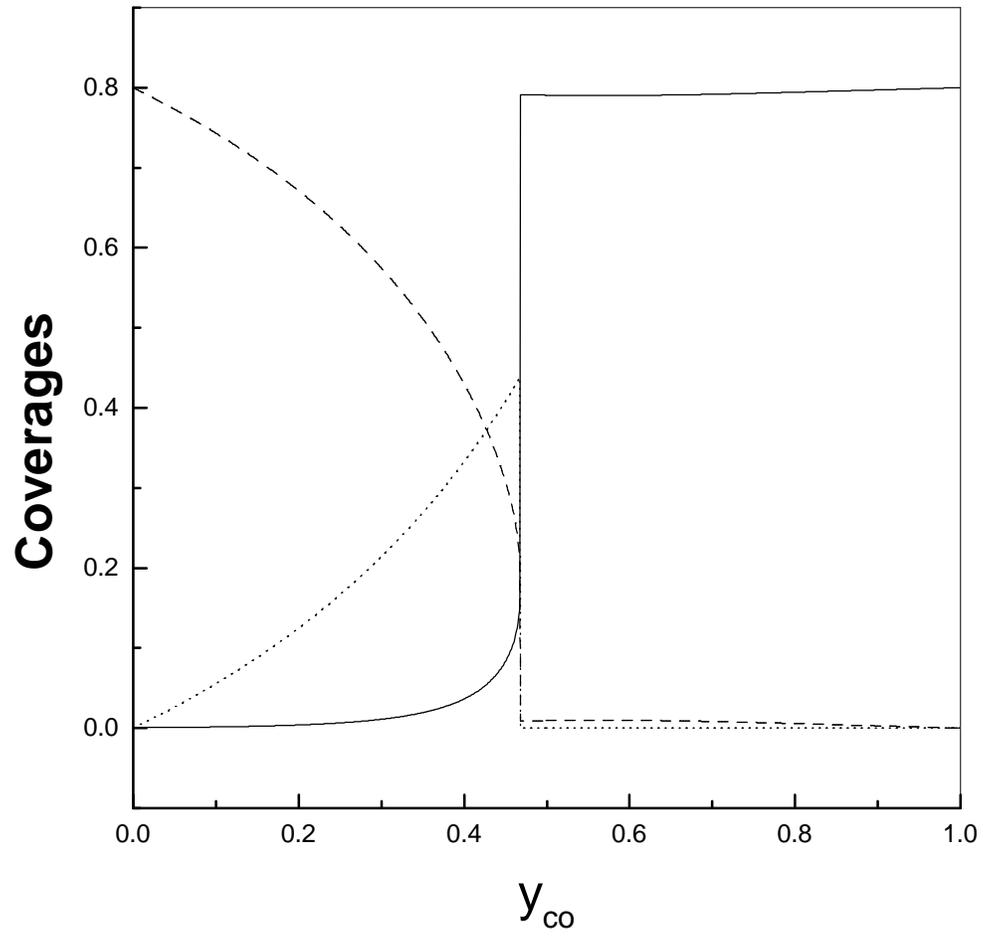
\begin{figure}
\caption{Coverages of ${\rm CO}$ (full line), ${\rm O}$ (dashed line) and empty sites (dotted line), as a function of the 
deposition rate of ${\rm CO}$. The concentration of inert sites is $n_d=0.2$. Coverages obtained through the site approximation. }
\end{figure}

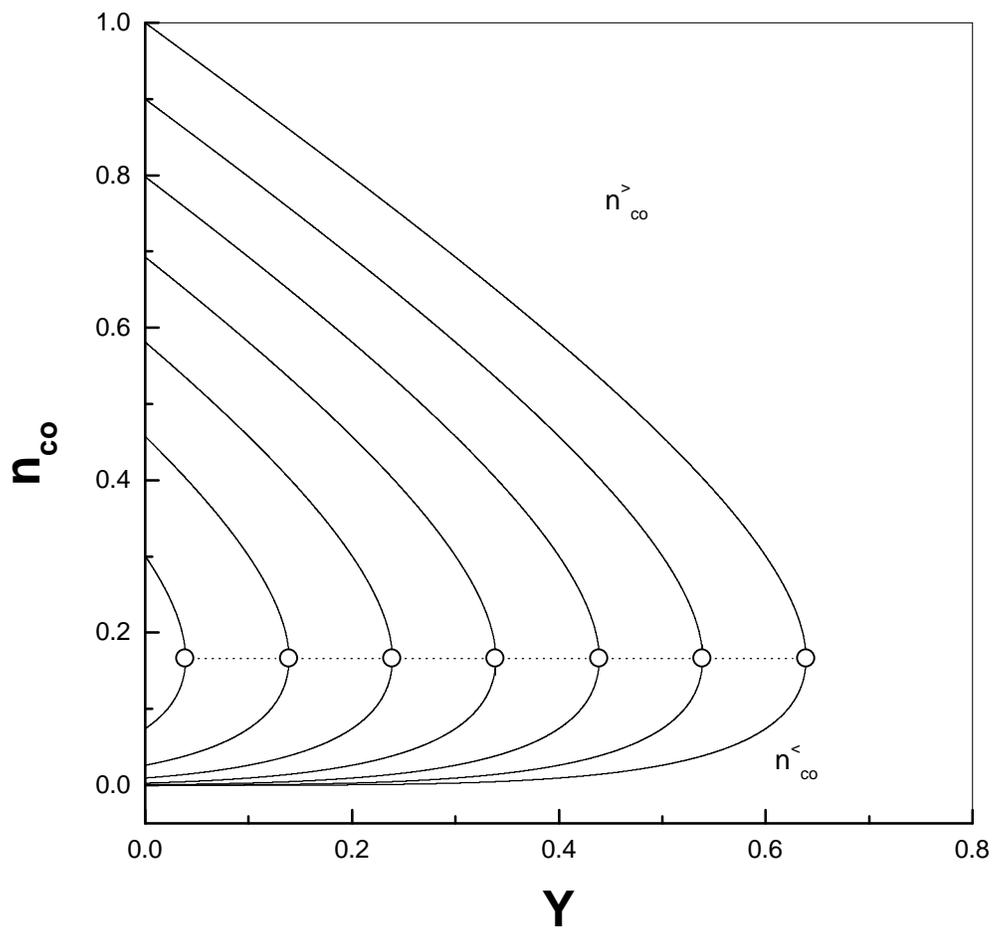
\begin{figure}
\caption{Stability curves in the site approximation for different values of concentration of inert sites. The upper ($n^{>}_{{\rm co}}$)  and lower ($n^{<}_{{\rm co}}$)  branches give the concentration of ${\rm CO}$, respectively, at the unstable and stable states. 
The open circles indicate the position of the spinodal points. From the outer to the  inner curves the concentration of inert sites is 0.0, 0.1, 0.2, 0.3, 0.4, 0.5 and 0.6. $Y=\frac{y_{{\rm co}}}{2(1-y_{{\rm co}})}$.}
\end{figure}

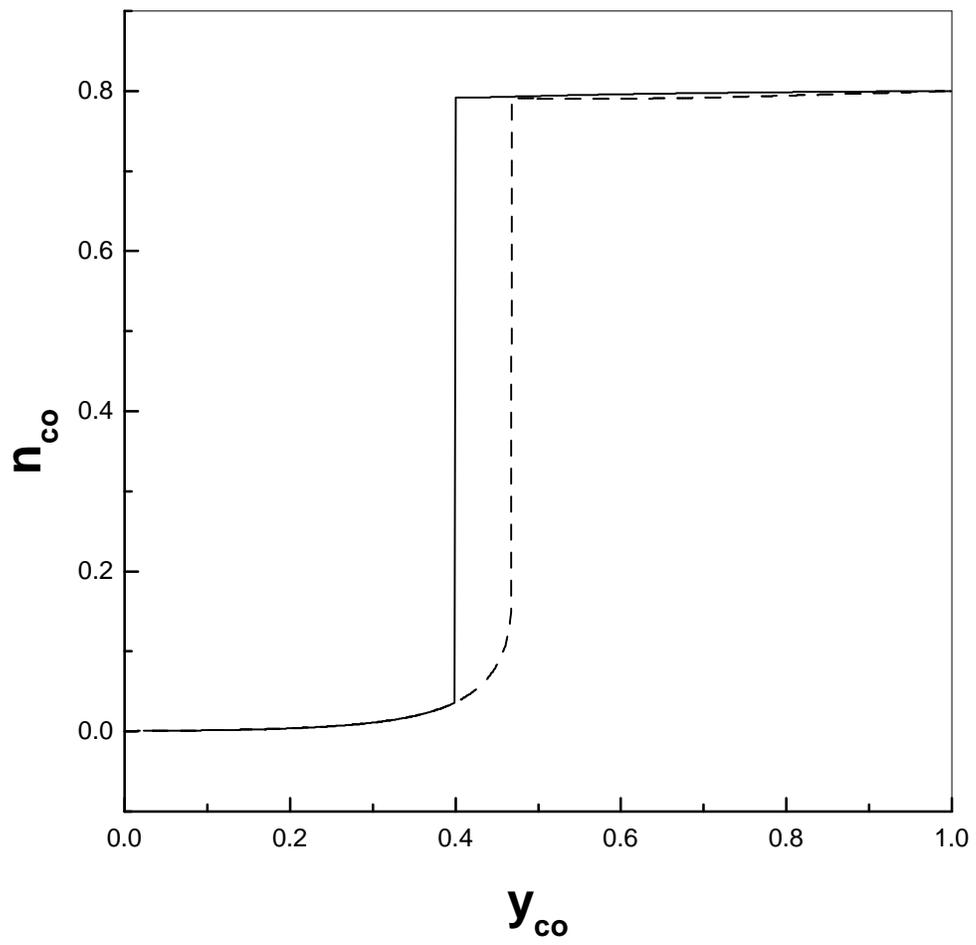
\begin{figure}
\caption{Plots of the concentration of ${\rm CO}$ ($n_{{\rm co}}$) versus $y_{{\rm co}}$ to locate the spinodal and transition points for the particular value  $n_d=0.2$. Spinodal (dashed line),  first-order transition (full line).}
\end{figure}

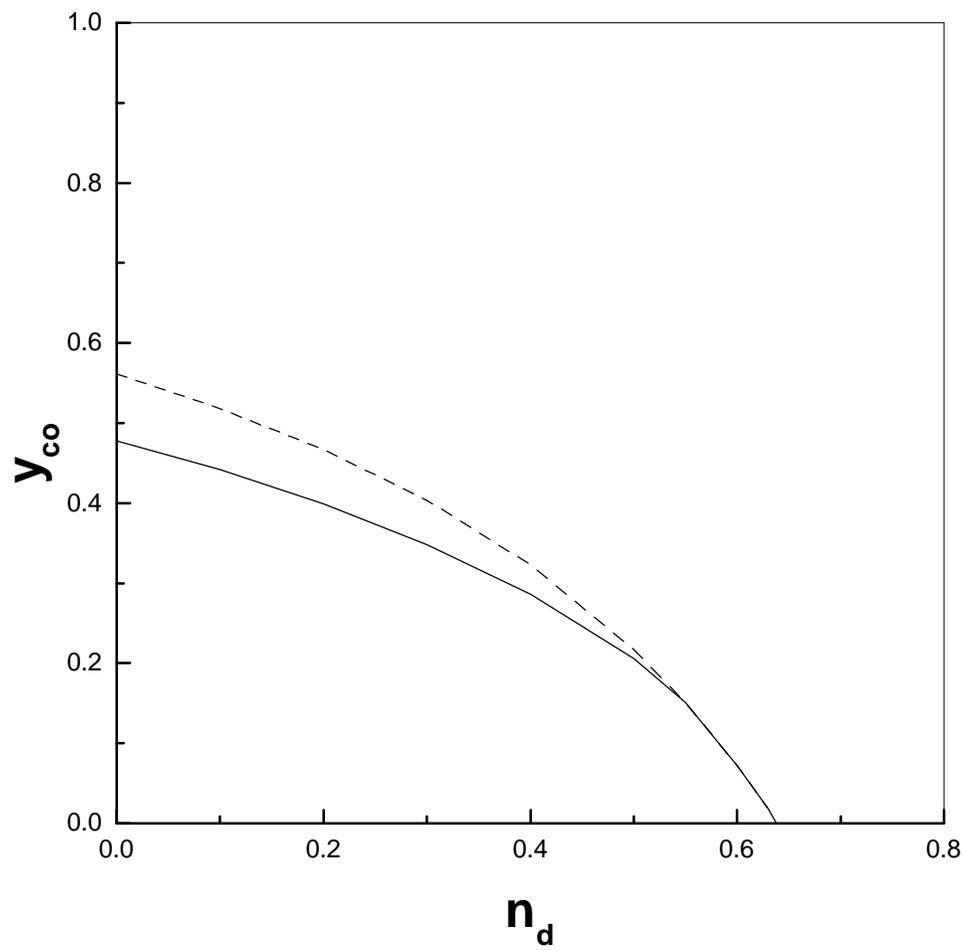
\begin{figure}
\caption{Behavior of the spinodal (dashed ) and first-order transition (full) lines  as a function of 
concentration of inert sites in the site  approximation.}
\end{figure}

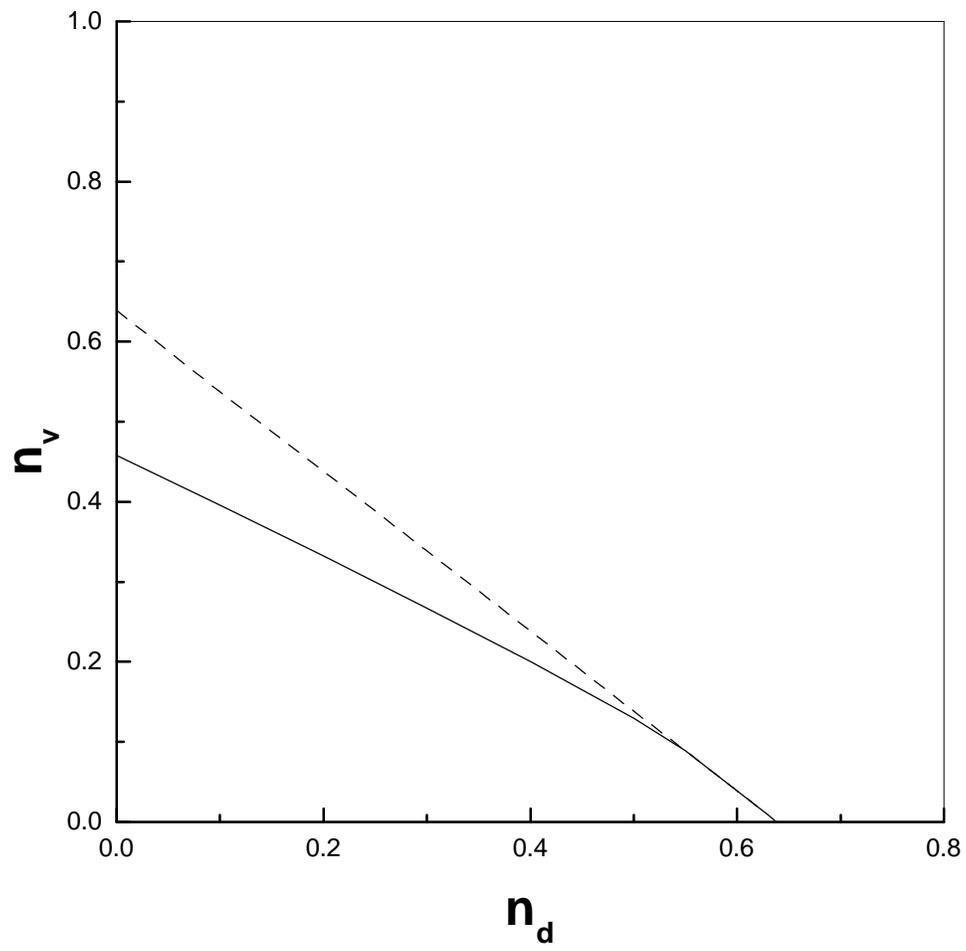
\begin{figure}
\caption{Concentration of empty sites at the spinodal (dashed line) and at the first-order transition 
(full line) as a function of the concentration of inert sites.}
\end{figure}

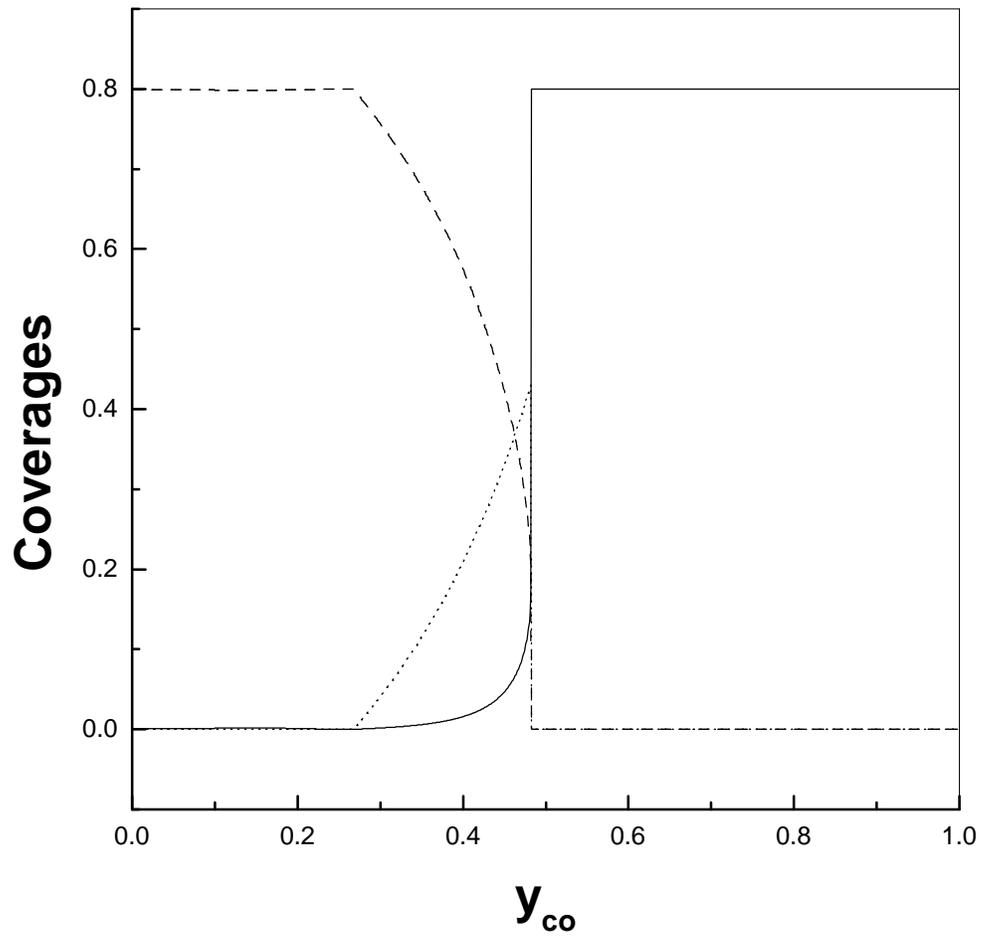
\begin{figure}
\caption{The same legend as in Figure 1, but  coverages obtained within pair approximation.}
\end{figure}

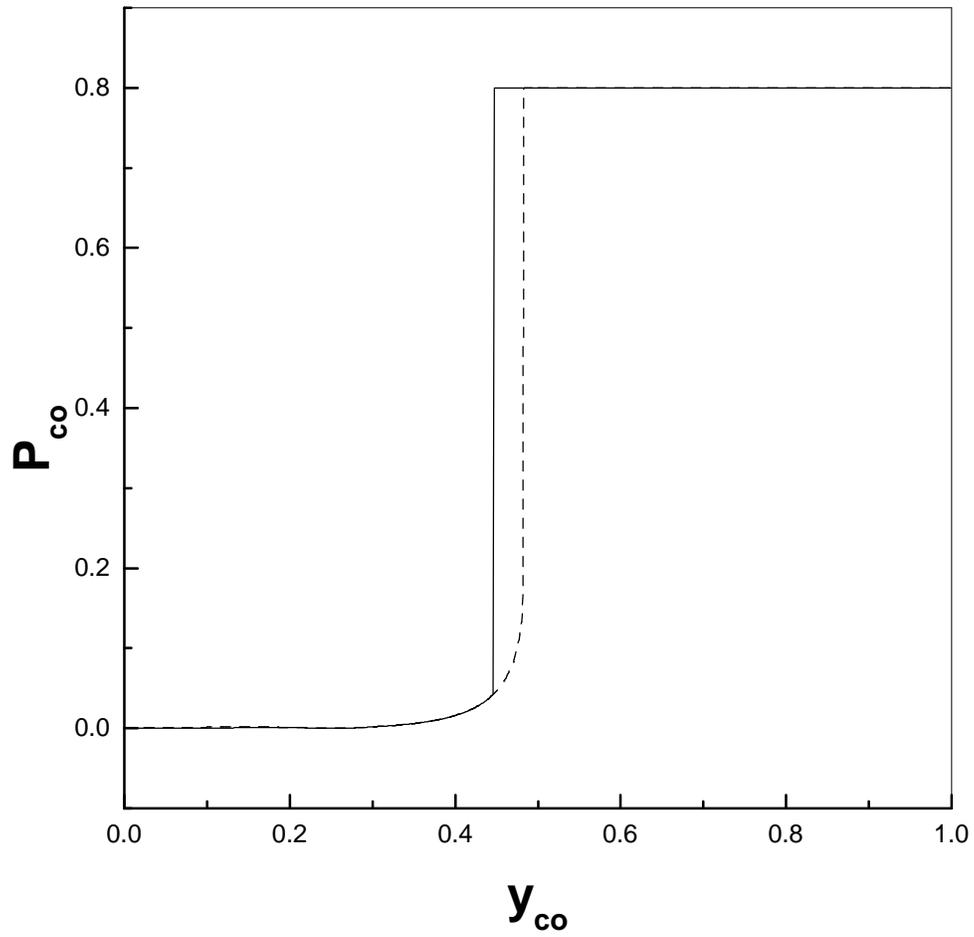
\begin{figure}
\caption{The same legend as in Fig. 3. Calculations performed in the pair approximation. $P_{{\rm co}}$ has the same meaning that $n_{{\rm co}}$.}
\end{figure}
    
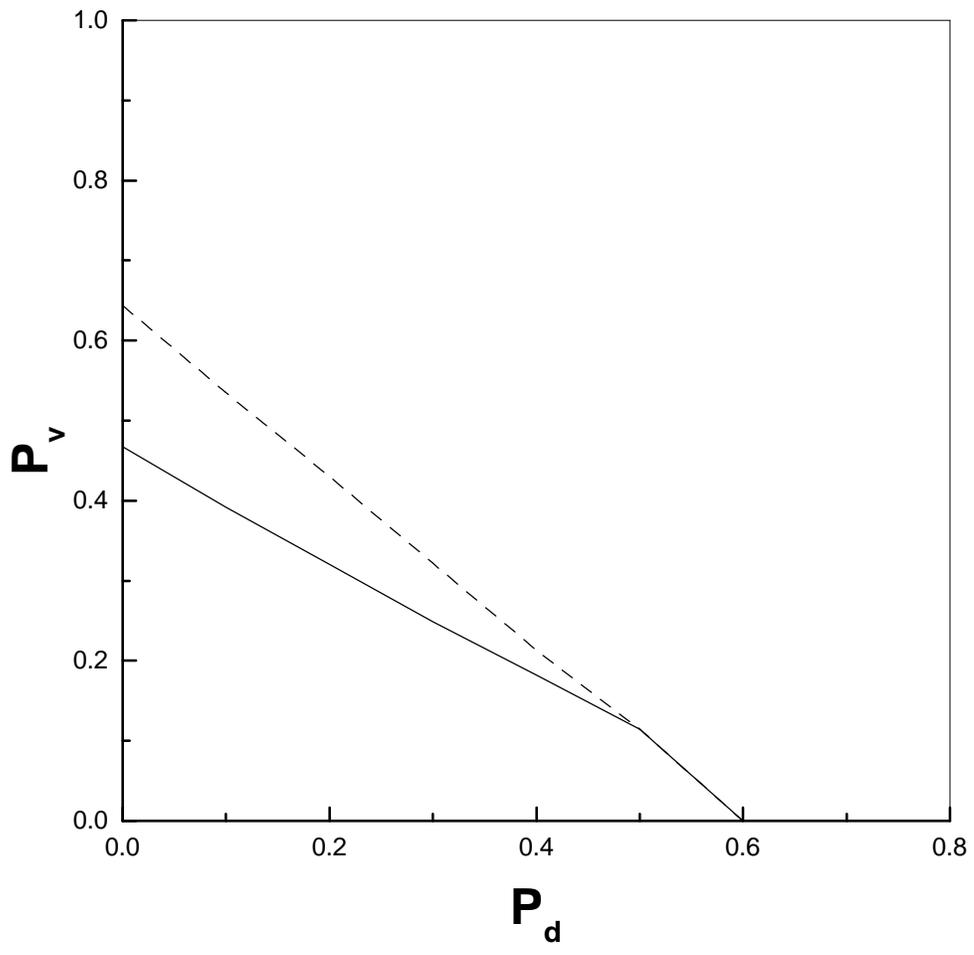
\begin{figure}
\caption{Concentration of vacant  sites at the spinodal (dashed line) and at the first-order transition 
(full line) as a function of the concentration of inert sites, in the pair approximation.}
\end{figure}

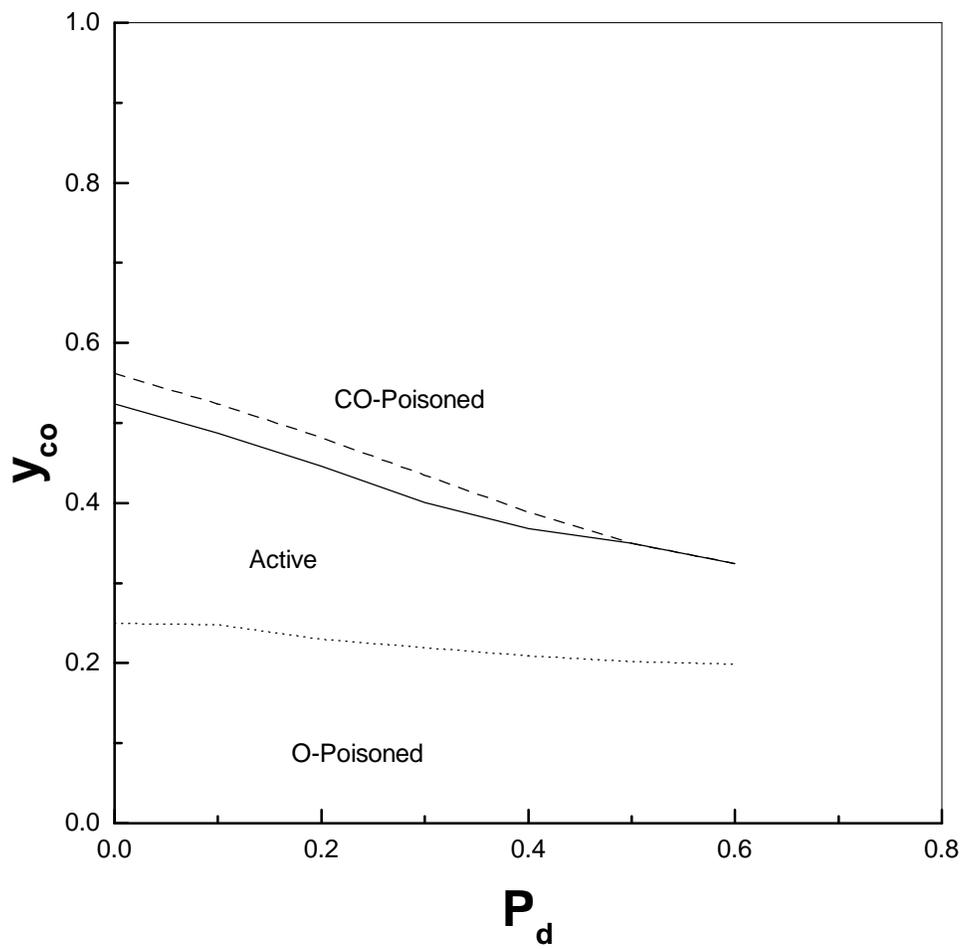
\begin{figure}
\caption{Phase diagram of the ZGB model with inert sites.  $P_d$ gives the concentration of inert sites, 
$y_{{\rm co}}$ is the probability that the  ${\rm CO}$ molecule hits  the surface. We have an active phase and a O-poisoned 
and CO-poisoned phases. The dashed line represents the spinodal points, the full line, the first-order 
transition points,  and  the dotted line gives  the continuous transition points. Phase diagram obtained in the pair approximation.}
\end{figure}

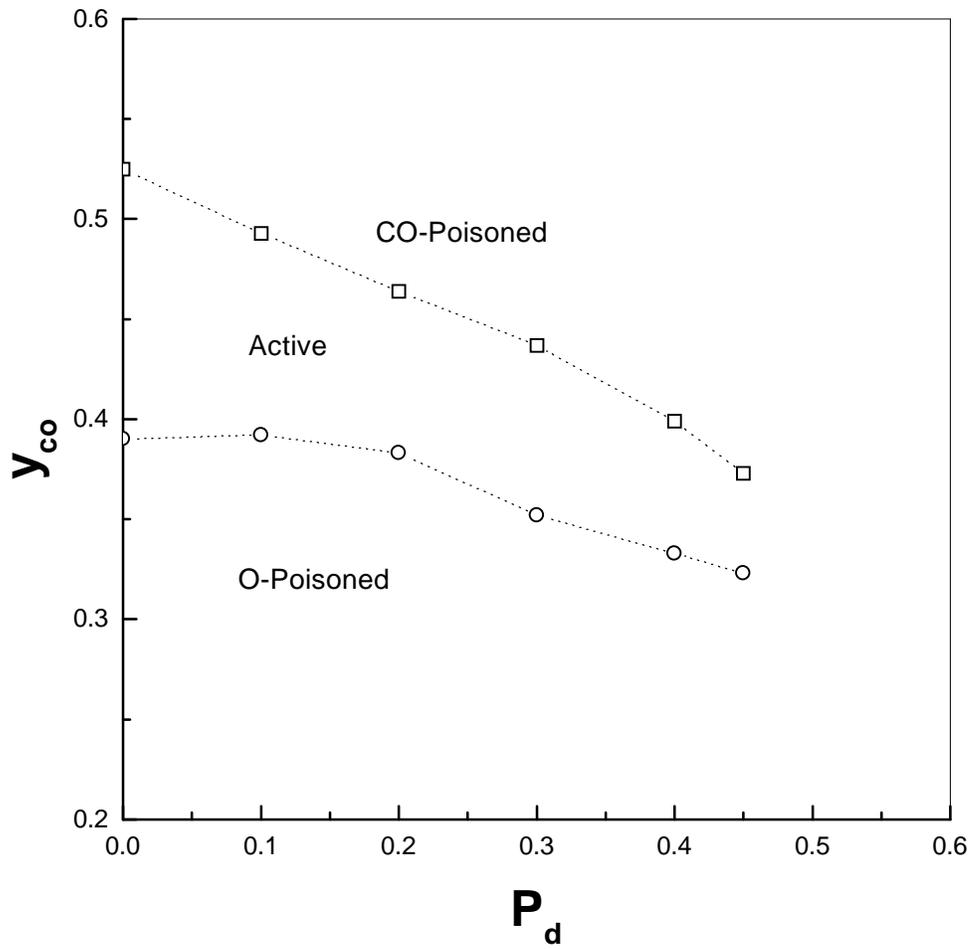
\begin{figure}
\caption{Phase diagram in the plane $P_d$ versus $y_{{\rm co}}$, obtained through Monte Carlo simulations showing the reactive window. The squares  
give  the transition between the reactive and CO-poisoned phases. The circles represent the continuous 
transitions between the active and O-poisoned phases. The lines are a guide to the eyes.}
\end{figure}

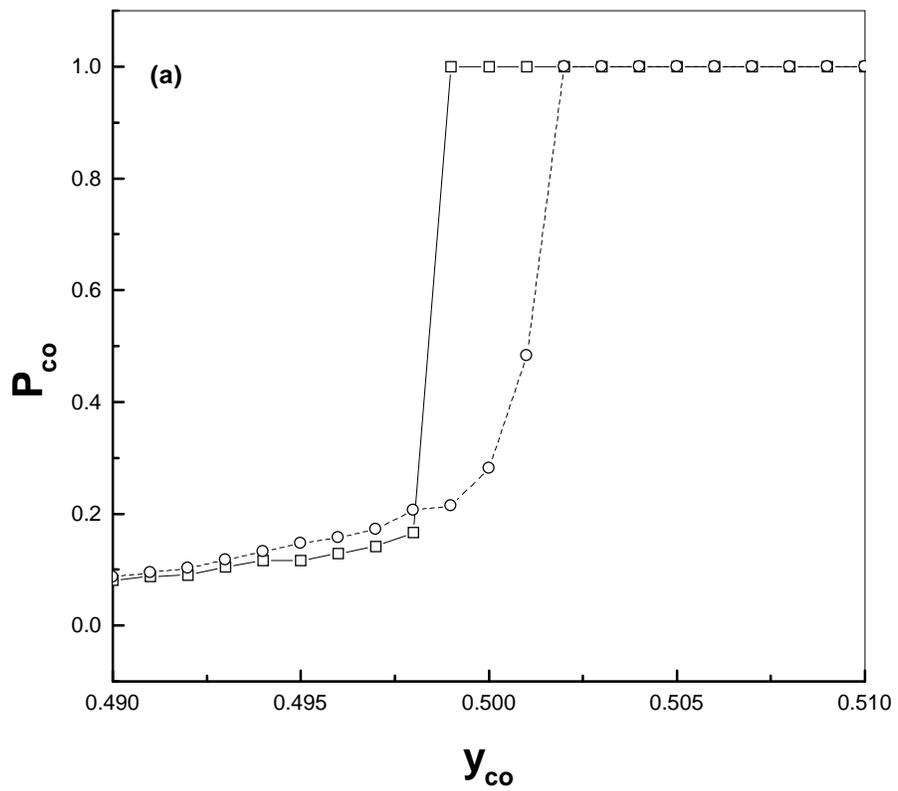
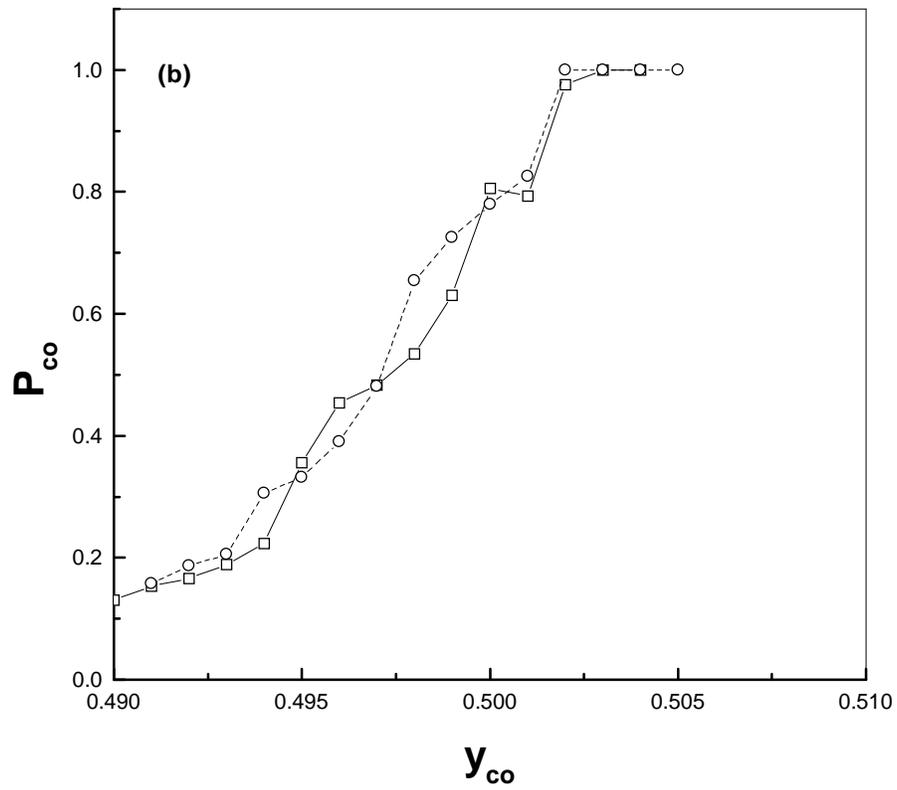
\begin{figure}
\caption{Hysteresis curves near the critical value of the concentration of inert sites. (a) $P_d=0.070$, (b) $P_d=0.080$.}
\end{figure}

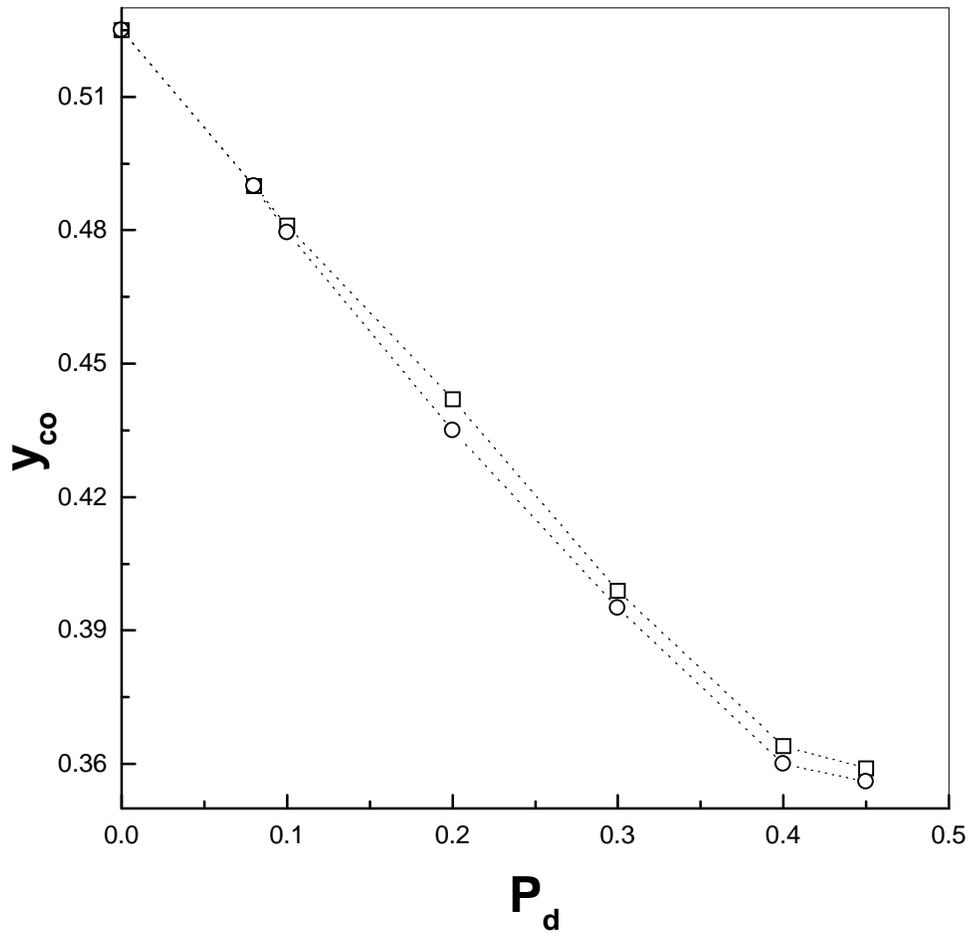
\begin{figure}
\caption{Plot of $P_d$ versus $y_{{\rm co}}$ showing the points where the production rate of ${\rm CO_2}$ is maximum (circles), and  the production  (squares) at the transition between the active and CO-poisoned phases. The lines serve as a guide to the eyes.}
\end{figure}
    
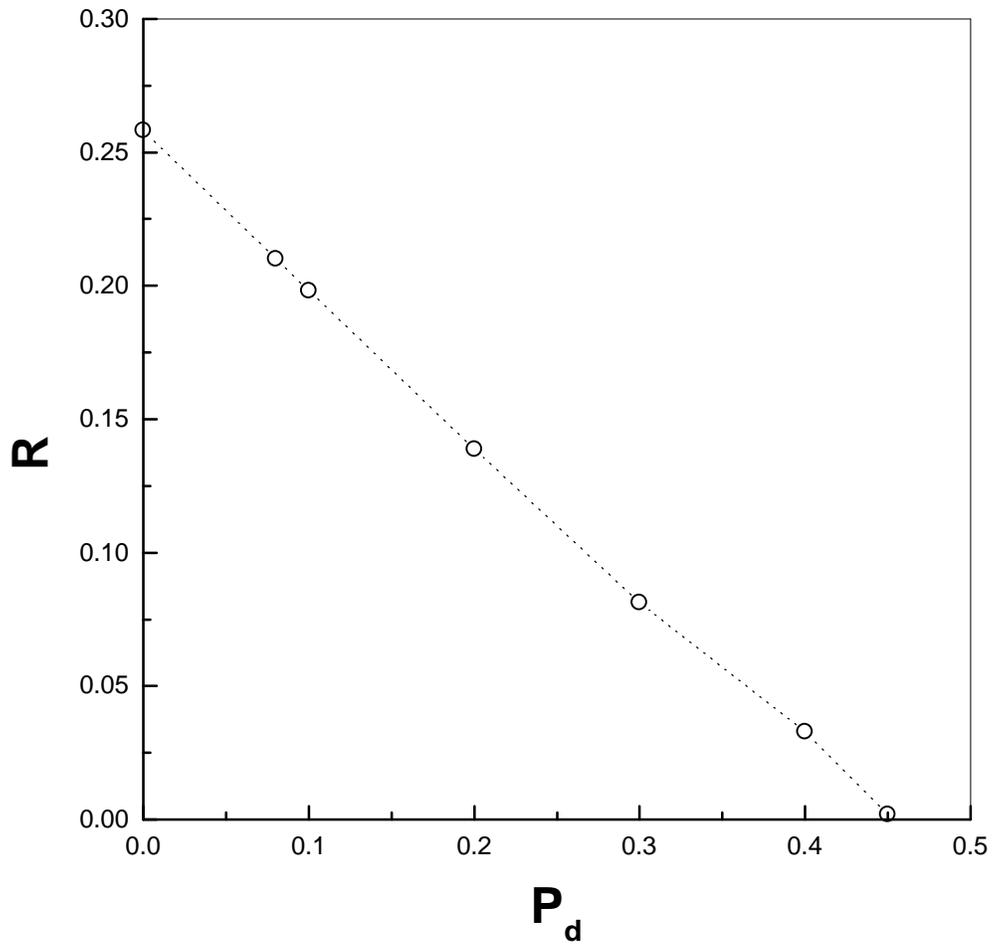
\begin{figure}
\caption{Maximum production rate  of ${\rm CO_2}$ molecules (R) as a function of the concentration of inert sites $P_d$.}
\end{figure}
              
\end{document}